\documentclass{PoS}
\usepackage[]{units}
\usepackage{amsmath}
\usepackage{hyphenat}
\def\apj{ApJ}%
\def\aap{A\&A}%
\def\mnras{MNRAS}%
\def\apjl{ApJ}%
\def\prd{Phys.~Rev.~D}%
\def\jcap{J. Cosmology Astropart. Phys.}%

\title{Gamma-ray cosmology and fundamental physics with TeV blazars: results from 20 years of observations }

\ShortTitle{Gamma-ray cosmology and fundamental physics with blazars}

\author{
\speaker{Jonathan Biteau} \& David A. Williams\\
\llap{}Santa Cruz Institute for Particle Physics and Department of Physics\\
University of California, Santa Cruz, USA\\
E-mail: \email{jbiteau@ucsc.edu}, \email{daw@ucsc.edu}}

\abstract{Gamma rays from TeV blazars have been detected by ground-based experiments for more than two decades. We have collected the most extensive set of archival spectra from these sources in order to constrain the processes affecting gamma-ray propagation on cosmological distances. We discuss our results on the diffuse photon field that populates universe, called the extragalactic background light, on the expansion rate of the Universe, and on fundamental physics in the form of axion-like particles and Lorentz-invariance violation. Specifically, we present a spectrum of the extragalactic background light from 0.26 to 105 microns constructed from the gamma-ray observations; we measure a value of the Hubble constant compatible with other estimates; and we constrain the energy scale at which Lorentz-invariance violation impacts gamma-ray absorption by the extragalactic background light to be larger than sixty percent of the Planck scale.}

\FullConference{The 34th International Cosmic Ray Conference,\\
		30 July- 6 August, 2015\\
		The Hague, The Netherlands}

\begin{document}

\section{Gamma-ray Transparency: A Convolution Product}

Gamma-ray cosmology is a line of research that aims to understand the propagation of TeV photons on cosmological distances, independently from the emission processes at play in the sources \cite{2013sf2aconf303B}. The main effect experienced by gamma rays on their journey is the production of electron-positron pairs in interactions with the extragalactic background light (EBL). The EBL is the second most intense cosmological background after the cosmic microwave background (CMB) and it encompasses the entire UV to far-infrared emission from stars and galaxies since the epoch of reionization. The H.E.S.S. and {\it Fermi}-LAT collaborations discovered the long-sought imprint of the EBL on the spectra of bright sources called blazars \cite{2013A&A...550A...4H,2012Sci...338.1190A} by jointly modeling the smooth intrinsic spectral curvature expected from radiation processes in the sources together with absorption by the EBL.  In \cite{2015arXiv150204166B}, we extend such analysis methods and investigate the archival gamma-ray spectra of blazars published over the past twenty years. 

The fraction of gamma rays, $\exp(-\tau)$, surviving interactions with EBL photons is a function of the gamma-ray optical depth $\tau(E_0,z)$, where $E_0$ is the gamma-ray energy measured in the lab frame and $z$ is the redshfit of the source \cite{1967PhRv..155.1404G}. The optical depth is the product of the distance between the emitter and the observer, the density of target photons, and the pair-production cross section. It thus results from a triple integration over the distance element, accounting for the cosmological model,\footnote{We assume here a concordance cosmological model with $H_0 = \unit[70]{km\ s^{-1}\ Mpc^{-1}}$, $\Omega_M=0.3$ and $\Omega_\Lambda=0.7$, but other cosmological models could in principle be taken into account within our formalism.} over the energy of the target photons, and over the incidence angle, with EBL photons assumed to be isotropically distributed in the comoving frame. We have shown in \cite{2015arXiv150204166B} that the integration over the incidence angle can be reduced analytically without any loss of generality. Going further requires an approximation, namely that the energy and redshift dependence of the EBL photon density can be decoupled, as initially proposed by \cite{1996ApJ...456..124M}. The optical depth can be then computed as a linear function of energy, multiplied by the convolution product of the EBL specific intensity at $z=0$, $\nu I_\nu$, and of a function that integrates the cross section over the line of sight, which we name the ``EBL kernel,'' $K_{z}$:

\begin{equation}
\tau(E_0,z_0) = \frac{3\pi\sigma_T}{H_0} \times \frac{E_0}{m_e^2 c^4}\times \nu I_\nu \otimes K_{z}\left(\ln\frac{E_0}{m_e c^2}\right)
\label{Eq:1}
\end{equation}
where $\sigma_T$ is the Thomson cross section, and $m_ec^2$ is the rest energy of an electron. The EBL intensity is here a function of $\ln(\epsilon_0/m_e c^2) = \ln(h\nu/m_e c^2)$,  with $\epsilon_0$ the EBL photon energy in the lab frame and $\nu$ the associated frequency, while the EBL kernel is a function of $\ln(E_0\epsilon_0/m_e^2 c^4)$, and the convolution integrates over the dependence on $\epsilon_0$ to leave a function of $\ln(E_0/m_e c^2)$.

\begin{figure}
\centering
\includegraphics[width=.72\textwidth]{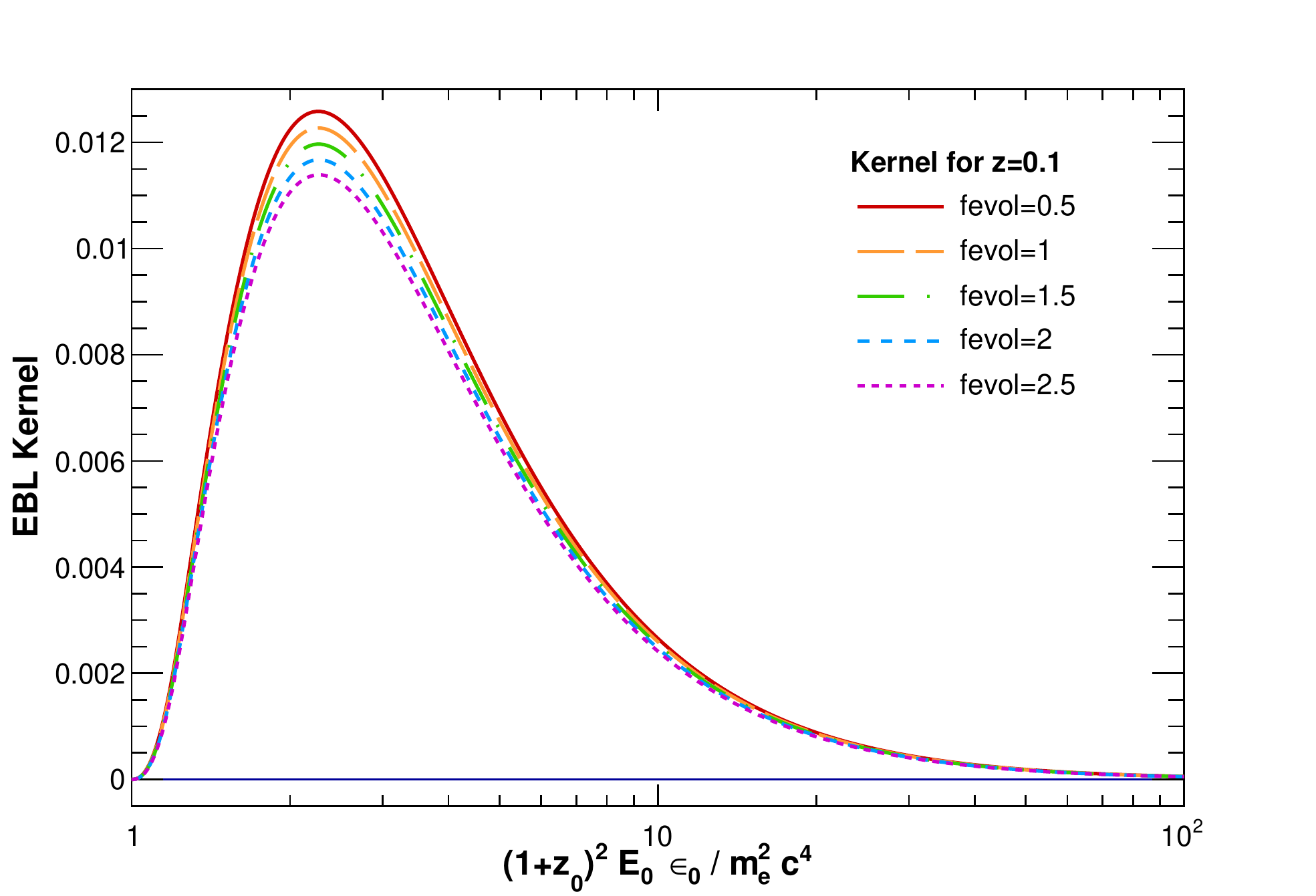}
\caption{EBL kernel for a source located at $z=0.1$ and various evolutions.}
\label{fig1}
\end{figure}

The EBL kernel, whose full expression can be found in \cite{2015arXiv150204166B}, is defined above the pair-production threshold, $
(1+z)^2 E_0 \epsilon_0/m_e^2 c^4 \geq 1$. The kernel shows a broad full width at half maximum, of about 5 in threshold units. The peak position and amplitude are increasing functions  of redshift, but do not strongly depend on the parameter $f_{\rm evol}$, which describes the evolution of the EBL with redshift, as shown in Fig.~\ref{fig1}. In \cite{2015arXiv150204166B}, we tuned the evolution parameter,  $f_{\rm evol}=1.7$, on the models of \cite{FR08} and \cite{G12}, resulting in absorption deviations to the models smaller than $15\%$ up to a redshift $z\sim0.8$. This is on the order of the systematic uncertainty on the flux quoted by ground-based gamma-ray instruments, justifying our approach. We assume a single evolution parameter for the entire EBL spectrum between $\sim0.1$ and $\sim\unit[100]{\mu m}$, but a natural extension of our approach could consist in developing a wavelength-dependent parametrization, possibly extending the applicability of this method beyond $z=1$. 

\section{Extracting the EBL Signal}

\subsection{Dataset and Deconvolution Method}

We have collected published gamma-ray spectra from Whipple, HEGRA, CAT, Tibet, ARGO-YBJ, TACTIC, HESS, MAGIC, and VERITAS, with observations spanning the period 1995-2013. Each spectrum is described with either a power-law, a log-parabola, or an exponential cut-off power-law model, whichever is preferred by the fit accounting for EBL absorption. Among the 106 spectra from 38 sources that we study, 86 spectra originate from known-redshift BL~Lacertae blazars, which we use for ``gamma-ray cosmology,'' i.e. reconstruction of the EBL, constraints on the Hubble constant, and limits on Lorentz-invariance violation. We also collected the spectral characteristics of the sources observed by {\it Fermi}-LAT at lower gamma-ray energies, when contemporaneous with ground-based observations. Assuming that both the low- and high-energy gamma rays originate from the same radiative component, which is the current understanding of BL~Lacertae emission, the {\it Fermi}-LAT observations below $\unit[100]{GeV}$ constrain the unabsorbed part of the spectrum, imposing a maximum hardness on the intrinsic emission at higher energies. 

For fitting purposes, we bin the EBL specific intensity in the wavelength range $\unit[0.26-105]{\mu m}$ in eight bins of width $\Delta \ln\lambda=0.75$. The maximum wavelength is determined by the pair-production threshold for the highest-energy gamma-ray spectral points, while the number of bins and minimum wavelength are limited by the gamma-ray statistics. The overall EBL spectrum is represented by a sum of Gaussians centered on each bin, with fixed widths and positions, and with free normalizations that are the fit parameters. The parameters of each intrinsic gamma-ray spectrum are left free to vary during the minimization procedure over the EBL parameters, following the approach of \cite{2013A&A...550A...4H}.

In addition to the quality of the fit of the gamma-ray spectra and to the maximum hardness limits, we can further constrain the EBL parameters using results from direct measurements and galaxy counts, which we largely extracted from \cite{2013APh....43..112D}. The former are sometimes prone to contamination from local foregrounds, such as zodiacal light, and are considered as upper limits in our minimization procedure. The latter only account for resolved populations of sources, and not for any truly diffuse component or undetected classes of objects such as primordial stars. We thus consider the results from galaxy counts as lower limits on the EBL intensity. More details on the data, deconvolution method, and minimization procedure can be found in \cite{2015arXiv150204166B}. For those gamma-ray spectra where the original authors have agreed, we plan on submitting the data points we have assembled to the ASDC SED builder and TeVCat,\footnote{http://tools.asdc.asi.it/, http://tevcat.uchicago.edu/} so they will be easily available to others.

\subsection{Best-fit EBL Spectrum}
\label{Sec:EBL}

\begin{figure}
\centering
\includegraphics[width=.72\textwidth]{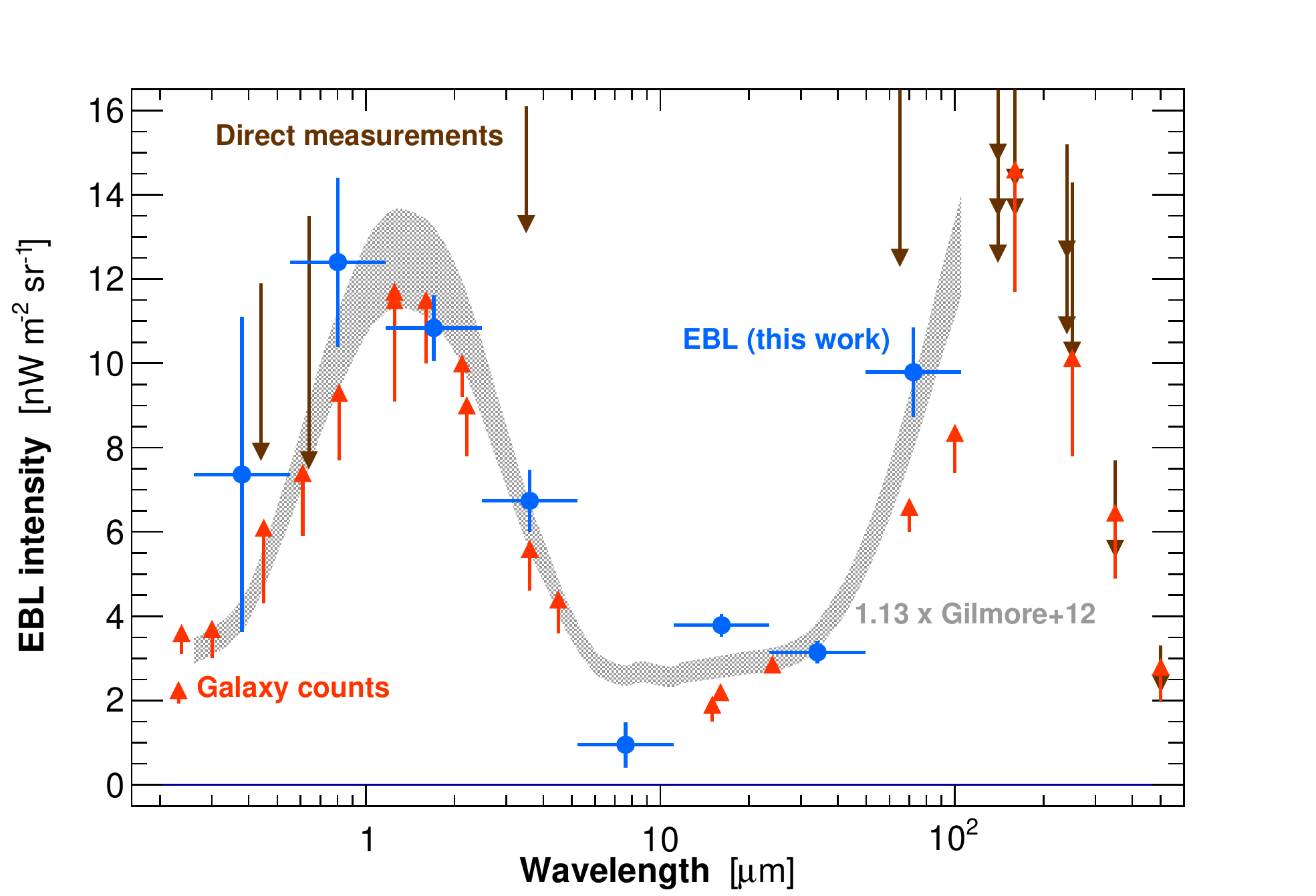}
\caption{Specific intensity of the EBL. Galaxy counts and direct measurements are extracted from the literature and represented by upward- and downward-going arrows, respectively. The model-independent result from this study based on gamma-ray observations is shown with blue points, while the model-dependent result based the EBL intensity of \cite{G12} is shown with a filled gray region ($\unit[1]{\sigma}$ confidence level).}
\label{fig2}
\end{figure}

The eight-point best-fit spectrum resulting from the 86 gamma-ray spectra and from local measurements is shown in Fig.~\ref{fig2}. Fixing the choice of intrinsic spectral model of each spectrum and leaving their parameters free, the best-fit EBL spectrum not accounting for local measurements is preferred by a likelihood ratio test at the $\unit[11]{\sigma}$ level with respect to no EBL. For comparison, we also show in Fig.~\ref{fig2} the result of a model-dependent approach where the normalization of the EBL spectrum modeled by \cite{G12} is left free to vary, reaching here a best-fit value of $1.13\pm 0.07$. We find that state-of-the-art EBL models \cite{FR08,G12,D11} represent the data equally well as our model-independent approach based on Gaussian sums, and we estimate the total brightness of the EBL in $\unit[0.1-1000]{\mu m}$ as $\unit[62\pm12]{nW\ m^{-2}\ sr^{-1}}$, or $\unit[6.5\pm1.2]{\%}$ of the brightness of the CMB. 

Mild discrepancies between the model-dependant and -independent approaches are observed around $\unit[10]{\mu m}$, which could possibly trace the signature of polycyclic aromatic hydrocarbons. We are in the process of investigating the impact of different reconstruction techniques, such as spline functions, to assess the robustness of these features. From the difference between gamma-ray constraints and galaxy counts, we can exclude any contribution to the EBL intensity larger than $\sim\unit[6]{nW\ m^{-2}\ sr^{-1}}$ from unresolved sources between 0.1 and $\unit[10]{\mu m}$, which rules out reionization scenarios involving miniquasars or primordial stars such as presented in \cite{2004MNRAS.351L..71C}.

\section{Beyond the Standard Models of Cosmology and Particle Physics}

\subsection{Hubble Constant}

The Hubble constant traces the expansion rate of the Universe. Measurements based on the cosmic distance ladder point to an expansion rate $H_0 = \unit[72.5\pm2.5]{km\ s^{-1}\ Mpc^{-1}}$, in $\sim\unit[2]{\sigma}$ tension with the latest CMB constraint from Planck, $H_0 = \unit[67.0\pm1.2]{km\ s^{-1}\ Mpc^{-1}}$ \cite{2014A&A...571A...1P}. Though this mild discrepancy might be merely of statistical origin, further tensions between local and distant measurements of the Hubble constant might reveal some limitations in the standard model of cosmology.

A number of authors, e.g. \cite{1994ApJ...423L...1S,2008MNRAS.389..919B, 2013ApJ...771L..34D}, proposed to use gamma-ray absorption as a probe of the Hubble constant. The first-order dependence of the optical depth on $H_0$ is shown in Eq.~\ref{Eq:1} and was used directly by \cite{2008MNRAS.389..919B} to set a lower limit on the Hubble constant. Similarly, by studying the ratio of the EBL reconstructed from gamma-ray observation and that inferred from galaxy counts, we reconstruct $H_0 =  \unit[88 \pm 8_{\rm stat} \pm 13_{\rm sys}\ ]{km\ s^{-1}\ Mpc^{-1}}$, for a fixed evolution parameter $f_{\rm evol}=1.7$. Due to its large uncertainties, our measurement is compatible at the $\unit[1.4]{\sigma}$ level with other state-of-the-art constraints, so that no hint for physics beyond the standard model of cosmology can be claimed at this stage. The authors of \cite{2013ApJ...771L..34D} notice that the density of photons could also depend on $H_0$ for a given set of galaxy observations. Within our formalism, this would translate into varying the evolution parameter as a function of the Hubble constant. Marginalizing over $0.5<f_{\rm evol}<2.5$, we reconstruct instead $H_0 =  \unit[88 \pm 13_{\rm stat} \pm 13_{\rm sys}\ ]{km\ s^{-1}\ Mpc^{-1}}$, which does not affect our conclusions.

\subsection{Lorentz-Invariance Violation}

The unification of quantum field theory, which describes physics at the particle scale, and of general relativity, relevant at cosmological scales or in extreme astrophysical conditions, is one of the most important quests of modern-day science. The search for a quantum-gravity theory has led to the development of new mathematical abstractions, such as loop quantum gravity or string theory (see, e.g., discussions in \cite{2001trqg.book.....S}), which have nonetheless sometimes been criticized for their lack of testable predictions. Phenomenologists have to a large extent deflected the criticism and bridged the gap between theory and experimental physics, by studying generic leading-order corrections to ``classical'' theories, and their testability in fields such as cosmology, particle physics, astroparticle physics, and gamma-ray astronomy \cite{2013LRR....16....5A}. 

One of these effects originates from a UV correction to the dispersion relation of particles, yielding  a norm of the momentum four-vector that is not a Lorentz invariant \cite{1999ApJ...518L..21K}. Assuming that the momentum four-vector is conserved between the initial and final states, one can compute a modified pair-production threshold. In the case of a subluminal first-order correction, the threshold condition becomes $(1+z)^2 E_0 \epsilon_0/m_e^2 c^4 \geq 1+(E_0/E_{\gamma, \rm LIV})^3$, with $E_{\gamma, \rm LIV} = [8E_{\rm QG}(m_ec^2)^2]^{1/3} \sim \unit[29.4]{TeV}\times(E_{\rm QG}/E_{\rm Planck})^{1/3}$ and where $E_{\rm QG}$ is the quantum gravity energy scale, assumed here to be on the order of the Planck energy scale $E_{\rm Planck} = \sqrt{\hbar c^5/G} = \unit[1.22 \times 10^{28}]{eV}$. Interestingly, despite a difference of fifteen orders of magnitude between gamma rays observed on Earth and the Planck energy scale, the pair-production threshold is already affected at the $\sim\unit[10]{\%}$ level around $\unit[15]{TeV}$.

\begin{figure}
\centering
\includegraphics[width=.72\textwidth]{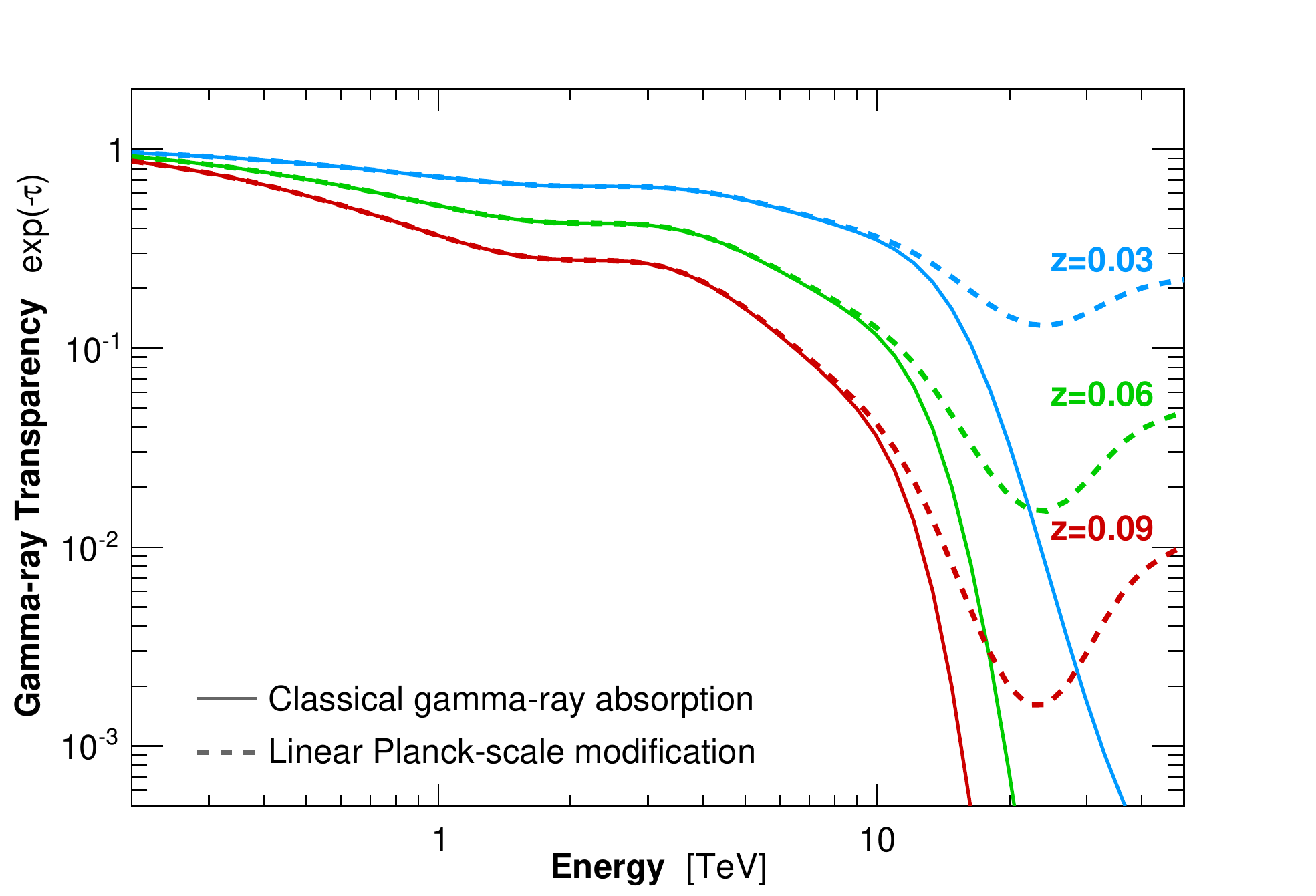}
\caption{Gamma-ray transparency as a function of energy in the classical case (solid lines) and with a modification of the pair-production threshold at the Planck scale (dashed lines).}
\label{fig3}
\end{figure}

Following the work of \cite{2008PhRvD..78l4010J}, we account for the modified threshold with a formalism similar to that presented in Eq.~\ref{Eq:1}. The net effect is an increased transparency to gamma-rays above $\unit[10]{TeV}$, largely independent of redshift, as shown in Fig.~\ref{fig3}. We constrain the effect by fitting to the 86 spectra a model including the quantum-gravity energy scale as a free parameter. The EBL parameters and the intrinsic parameters are left free during the minimization procedure. The likelihood profile as a function of $E_{\rm QG}$ shows a maximum around the Planck energy scale, though only significant at the $\unit[2.4]{\sigma}$ level. We set the first quantitative limit on the effect, accounting for the uncertainties on the EBL and on the intrinsic emission of BL~Lac objects, as $E_{\rm QG}>0.6\times E_{\rm Planck}$ at the $\unit[95]{\%}$ confidence level. Future observations with instruments sensitive to higher-energy gamma rays, such as CTA \cite{2013APh....43....3A}, HAWC \cite{2013APh....50...26A}, and LHAASO \cite{2010ChPhC..34..249C}, will be able to rule out or unveil a modification of the pair-production threshold at the Planck scale.   

\subsection{Signatures of Ultra-High Energy Cosmic Rays and Axion-Like Particles}

The study of Lorentz-invariance violation is a natural, one-parameter extension of the standard picture, in which gamma rays are emitted at the source and pair produce on the photons emitted by stars and galaxies since the end of the dark ages. More complex extensions of the standard picture have been proposed. A first one, based on classical physics only, relies on the acceleration and escape of a significant number of ultra-high energy hadrons from the sources. These hadrons would then produce neutral pions in interactions with the EBL, generating the VHE radiation observed on Earth \cite{2010APh....33...81E}. A second one involves the conversion of gamma rays to hypothetical axion-like particles in magnetic fields of the blazar jet, galaxy, or galaxy cluster (e.g. \cite{Hooper}). These axion-like particles then propagate freely through the EBL and potentially convert back to gamma rays in the magnetic field of the Milky Way. Both effects have been claimed to explain some features of the experimental data, such as possibly harder-than-expected spectra at the highest energies or optical depths (e.g. \cite{2012JCAP...02..033H}).

The parameters of these models vary from source to source (e.g. strength of the magnetic fields, fraction of hadrons escaping the source, etc), so that detailed constraints require studies beyond the work presented in \cite{2015arXiv150204166B}. We can nonetheless question the need for such scenarios by investigating the match between the observed gamma-ray spectra and EBL absorption alone. Note that for sources with underconstrained redshifts, we fix the distances to their best estimates based on the various constraints found in the literature. Based on the results presented in Sec.~\ref{Sec:EBL}, we reach the following conclusions: I) The 106 spectra we studied are well matched by gamma-ray absorption, with $\chi^2$ probabilities larger than $\unit[9]{\%}$ for all of them. II) We do not find any tension with the hardness limit imposed by the {\it Fermi}-LAT observations. III) The level of EBL reconstructed from gamma-ray observations only is in good agreement with constraints from direct measurements and galaxy counts. IV) We do not find any correlation between the fit residuals and the energy or the optical depth, at odds with the conclusions of \cite{2012JCAP...02..033H}. We conclude that the standard picture is, at this stage, sufficient to explain gamma-ray observations of blazars, without the need for scenarios involving axion-like particles or ultra-high energy cosmic rays. We encourage further constraints on these scenarios using state-of-the-art EBL models and a large unbiased sample of gamma-ray spectra.\\  

We gratefully acknowledge support from the U.S. National Science Foundation award PHY-1229792. We would also like to thank Matthieu Bethermin, Michael Dine, Alberto Dominguez, Steve Fegan, Amy Furniss, Joel Primack, and Stefano Profumo for helpful discussions.

\providecommand{\href}[2]{#2}\begingroup\raggedright\endgroup

\end{document}